# Integrated Electro-Optic Absorption Modulator for Silicon Nitride Platform

Evgeniy S. Lotkov[1,2], Alexander S. Baburin[1,2], Ali S. Amiraslanov[1], Evgeniy Chubchev[2], Alexander Dorofeenko[2,3], Evgeniy Andrianov[2,3], Ilya A. Ryzhikov[2,3], Evgeny S. Sergeev[1,2], Kirill Buzaverov[1,2], Sergei S. Avdeev[1,2], Alex Kramarenko[1,2], Sergei Bukatin[1,2], Victor I. Polozov[2,3], Olga S. Sorokina[1,2], Yuri V. Panfilov[1], Daria P. Kulikova[2,4], Alexander V. Baryshev[2] and Ilya A. Rodionov[1,2,*]

[1]FMN Laboratory, Bauman Moscow State Technical University, Moscow, Russia
[2]Dukhov Automatics Research Institute, (VNIIA), Moscow, Russia
[3]Institute for Theoretical and Applied Electromagnetics RAS, Moscow, Russia
[4]Faculty of Physics, Lomonosov Moscow State University, Moscow, Russia
*irodionov@bmstu.ru

**ABSTRACT**

Silicon nitride (SiN) is currently the most prominent platform for photonics at visible and near-IR wavelength bandwidth. However, realizing fast electro-optic (EO) modulators, the key components of any integrated optics platform, remains challenging in SiN. Recently, transparent conductive oxides (TCO) have emerged as a promising platform for photonic integrated circuits. Here we make an important step towards exceeding possibilities of both platforms, reporting for the first-time high-speed ITO electro-optic modulators based on silicon nitride waveguides. The insertion losses of 5.7 dB and bandwidth of about 1 GHz are shown for 300 nm-thickness SiN waveguide platform with 9.3-um-length hybrid waveguide. The fabrication process of devices requires only standard clean room tools, is repeatable and compatible with the CMOS technology. Simulation results of optimized device designs indicate that further improvement is possible and offer promising opportunities towards silicon nitride photonic computation platforms based on ITO.

## Introduction

Today, SiN photonics is definitely one of the main integrated photonic platforms. A key contributor to its success is the SiN high transparency window, which together with its low losses opens roads to wide range of applications [1-4]. However, realizing the key components of any integrated optics platform – fast electro-optic modulation – remains challenging in SiN. One of the promising EO materials for SiN based modulators is Indium Tin Oxide (ITO). Transparent conductive oxides including ITO are adopted in high-tech industry fields such as in touchscreen displays of smartphones or contacts for solar cells. Recently, ITO has been explored for electro-optic (EO) modulation using its free-carrier dispersive effect enabling strong index modulation [5-8]. One can certainly assert that there are applications where established EO materials such as lithium niobate (LiNbO$_3$, LN), especially bonded to SiN, has particular modulation advantages. However, photonic computations demand very dense integration of over ~10$^4$ optical components, where device footprint of even 100 μm$^2$ starts to impact performance. For such applications, ITO provides synergistic benefits when monolithically and heterogeneously integrated with low-cost SiN PICs. It is also worthwhile to mention here that, the demonstrated capabilities of ITO as a strong nonlinear material can designate desirability as one can combine many different functionalities on the same chip with the same active material [9, 10]. We have previously studied ITO properties tuning and have shown that it can be selectively adapted by process conditions for operation in either an n-dominant or α-dominant region defined by the level of the carrier concentration [11].

We demonstrate modulators based on the TM-polarized modes that provide a plasmonic enhancement of the field interaction with an electrooptical ITO film. The obtained results pave the way for a comprehensive platform of heterogeneous integration of ITO-based electro-optic devices into SiN PICs.

## Results and Discussion

**Design and fabrication.** The device consists of a photonic waveguide and active element section in a form of hybrid waveguide.

The photonic waveguide is a 300 nm × 2900 nm Si$_3$N$_4$ ridge on SiO$_2$ substrate. The waveguide is single-mode in terms of the TM waves. Although TE modes exist at the same frequencies, they are not excited in our experiments.

The active element is an absorption modulator with a 9.3-μm-length hybrid waveguide, in which the TM mode is coupled to surface plasmon-polariton (SPP) propagating along the metal-oxide interface (Fig. 1a), as suggested in [12] for Si waveguides (further – TM hybrid modulator). The application of a drive voltage places the capacitor formed by the Al and ITO films into the states of accumulation or depletion, thus changing the ITO electrode carrier concentration and hence its optical complex index due to the free-carrier plasma dispersion effect. In order to extract relevant parameters including the effective indices (real and imaginary parts, $n_{eff}$ and $k_{eff}$) we perform FEM eigenmode analysis for our structure. The corresponding mode profiles of the photonic ($Si_3N_4$) and hybrid ($Si_3N_4/ITO/Al_2O_3/Al$) waveguides are displayed as insets to the plot for the TM mode $n_{eff}$ dependence on SiN waveguide height (Fig. 1b). It should be noted that the fundamental TM mode does not exist when the SiN waveguide in the top air cladding has a height of less than 300 nm. Then, we compare the efficiency parameters over the 20 nm-thick ITO films electron concentration in the considered design at applying voltage from -7 to +7 V (Fig. 1c).

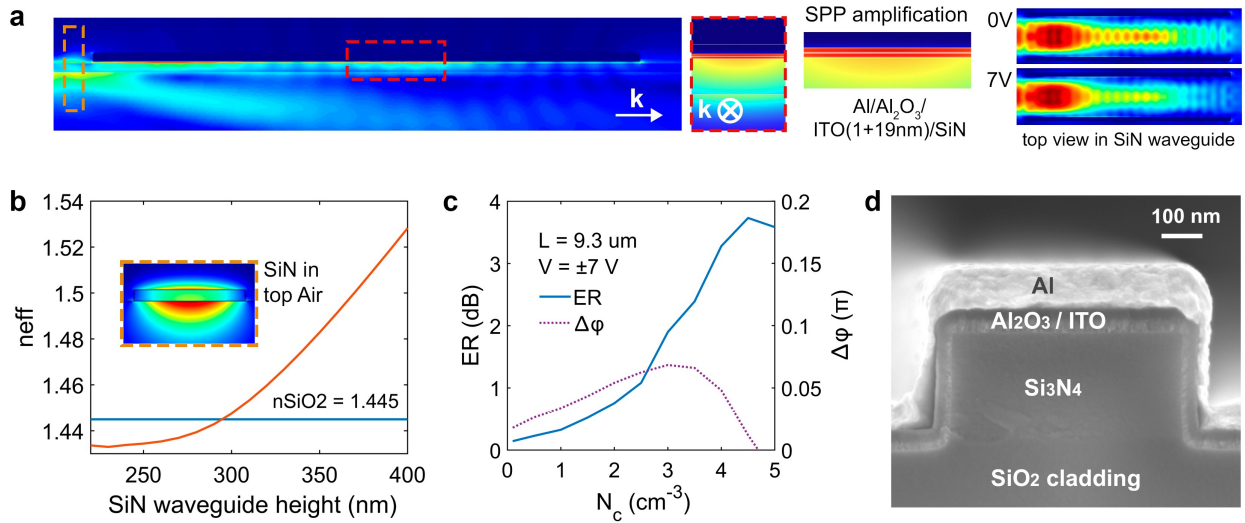

**Fig. 1** TM hybrid modulator on SiN platform: (a) 3D simulation of the $TM_{00}$ mode coupling and propagation in the hybrid waveguide; (b) dependence of $n_{eff}$ of the $TM_{00}$ mode on the SiN waveguide thickness at w = 2900 nm (the mode profiles of the photonic and hybrid waveguides are displayed as insets); (c) simulation of ER and phase change dependences on the initial ITO carrier concentration at ±7 V; (d) cross section SEM image of the fabricated device.

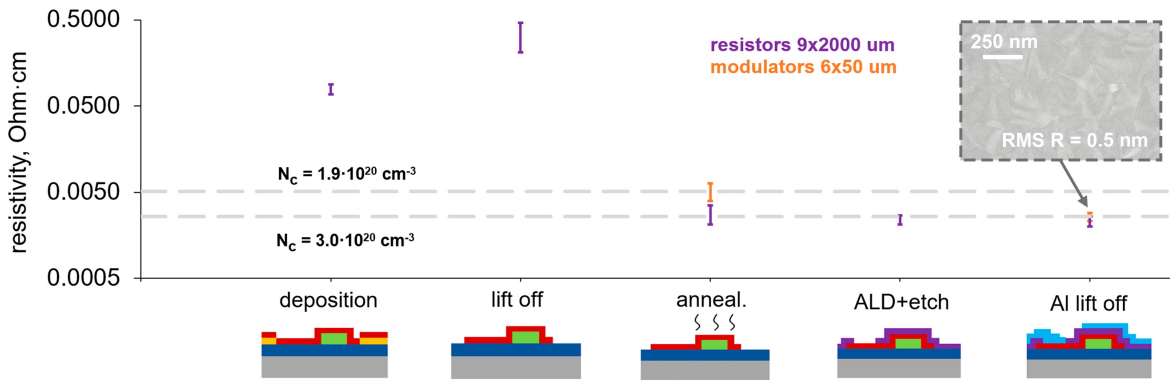

**Fig. 2** Evolution of the ITO elements resistivity through the fabrication cycle (ITO SEM image on the final study is displayed as insert).

The SiN single mode waveguides were fabricated (see Methods). The subsequent processing for the active device section includes deposition of the ITO film on the fabricated device using ion beam assisted evaporation (IBAD) and a top plasmonic metal (Ti/Al) stack on the top of the hybrid waveguide separated by a 20 nm-thick dielectric layer for TM hybrid modulator. It should be noted that IBAD provides high-quality ITO as it yields dense crystalline pinhole-free films with high uniformity as analyzed by us earlier [11]. With respect to nanophotonic device fabrication, as required here, IBAD enables precise control of material properties such as microstructure, non-stoichiometry, morphology, crystallinity. We use a relatively high-k dielectric, $Al_2O_3$, for the gate oxide fabricated by atomic layer deposition (ALD) since the modulation efficiency improves with enhanced electrostatic characteristics [13]. Fig. 1d shows the reference cut of the final modulator stack SEM image. After a lift-off process the initial carrier concentration of ITO elements was $N_{c\_lift-off} = 1.9 \cdot 10^{20}$ cm$^{-3}$ measured by electrical tests and ellipsometry, but due to the fabrication process, in the final stage we obtained $N_{c\_final} = 3.0 \cdot 10^{20}$ cm$^{-3}$ (Fig. 2). A more detailed fabrication process description can be seen in Methods.

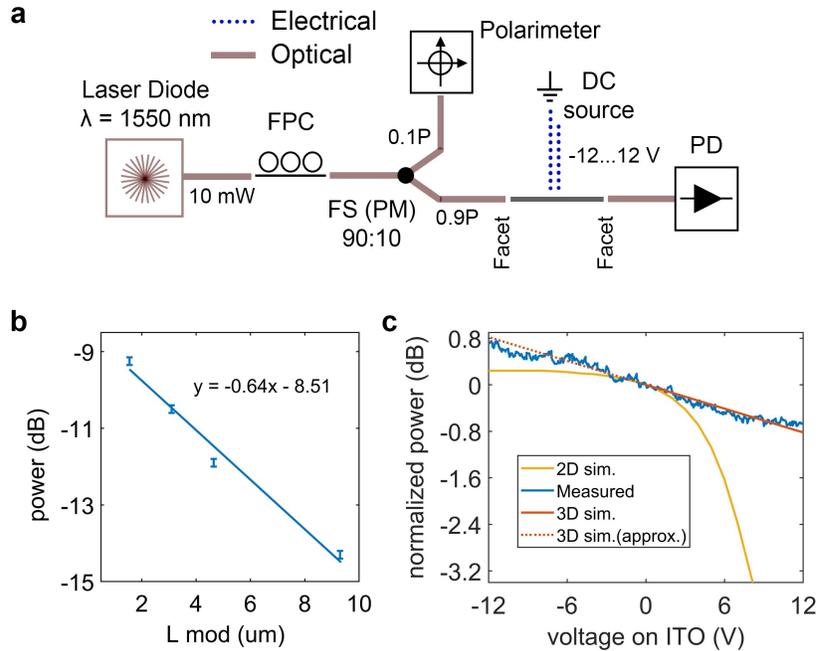

**Fig. 3** Insertion loss and electro-optic efficiency characterization: (a) measurement setup; (b) insertion losses of the modulators with different lengths; (d) DC amplitude modulation measurements and simulation results for 9.3 um-long device.

**On-chip insertion loss.** To characterize the optical loss by the cutback method, TM polarized light (QDFB LD-1550-50N laser diode, Qphotonics) was coupled to the chip by the special fabricated facet [14] using single mode fiber (SM15, coupled loss with facet was measured as 3.5 dB). The light intensity was measured with a photodetector (S154C, Thorlabs). Fig. 3a demonstrates the measurement setup.

The insertion loss of the TM hybrid modulator was 6.3 dB (on-chip IL = 14.6 dB). Losses for the photonic-plasmonic junction were 0.6 dB (simulations show that these losses could be 2.2 dB per junction). The plasmonic propagation losses were 0.64 dB/μm (Fig. 3b). Propagation losses of the Si$_3$N$_4$ waveguides with different lengths at TM polarization were 1.29 dB/cm.

As shown by simulation results, the insertion loss of the 9.3 μm-length TM hybrid modulator at the initial ITO carrier concentration of $3.0 \cdot 10^{20}$ cm$^{-3}$ are 9.4 dB. The difference in losses between the fabricated device and the simulation could be explained by numerical overestimation of photonic-plasmonic coupling. In experiment, the TM mode could be not properly coupled with hybrid waveguide (1.2 dB experimental junction losses versus 4.4 dB the simulation one). It can be further improved by increasing of Si$_3$N$_4$ waveguide thickness that provides the better coupling.

**Electro-optic efficiency.** The electrically induced amplitude modulation in absorption design was measured. We applied voltage to the modulator using DC probes (Keysight 2901B) and measured the intensity drop at a fixed wavelength λ = 1550 nm using a laser diode and photodetector (Fig. 3a). Fig. 3c shows the transfer function of the insertion loss versus applied voltage. Extinction ratio of 1.5 dB (± 12 V) and 0.85 dB (± 7 V) are achieved.

From 3D simulations of 9.3 μm modulator ER was calculated to be 3.7 dB (at ±12 V) and 1.9 dB (at ± 7 V). The discrepancy may be caused by the same reason why the modulator has losses less than calculated, namely by the photonic-plasmonic junction, which can occupy a larger part of the plasmonic element. This results in a smaller effective length in which interaction with the hybrid mode occurs.

**High-speed characterization.** Fig. 4a illustrate the schematic of the experimental testing setup. Electric microwave was induced by vector network analyzer (R&S ZNB 20) to measure s-parameters. To minimize the parasitic reflections of microwaves supplied to the modulator, we optimized the shapes and fabricated the coplanar metal electrodes of the one-port and two-port versions (Fig. 4b shows the one-port one). The one-port version shows reflections greater than -5 dB over the entire frequency range and requires matching the impedance of the MOS capacitor to a specific fixed frequency, in contrast to the two-port version, which shows reflections at -15 dB level (Fig. 4c). Next, the normalized frequency electro-optical response of the modulator was measured (Fig. 4d). Based on the geometry derived from SEM images (Fig. 4b), we calculated a device capacitance of 162 fF (measured value is 400 fF) and a resistance of 717 Ω originating mainly from the ITO electrode (measured value is 1140 Ω). The simulated electrical 3dB bandwidth should therefore be 1.36 GHz. Fitting the response at 1550 nm indicate 3dB bandwidth limitation of 1.05 GHz and 6dB bandwidth limitation of 2.3 GHz. With geometry optimization, the real frequency response of the device is expected to be significantly beyond 260 GHz [7].

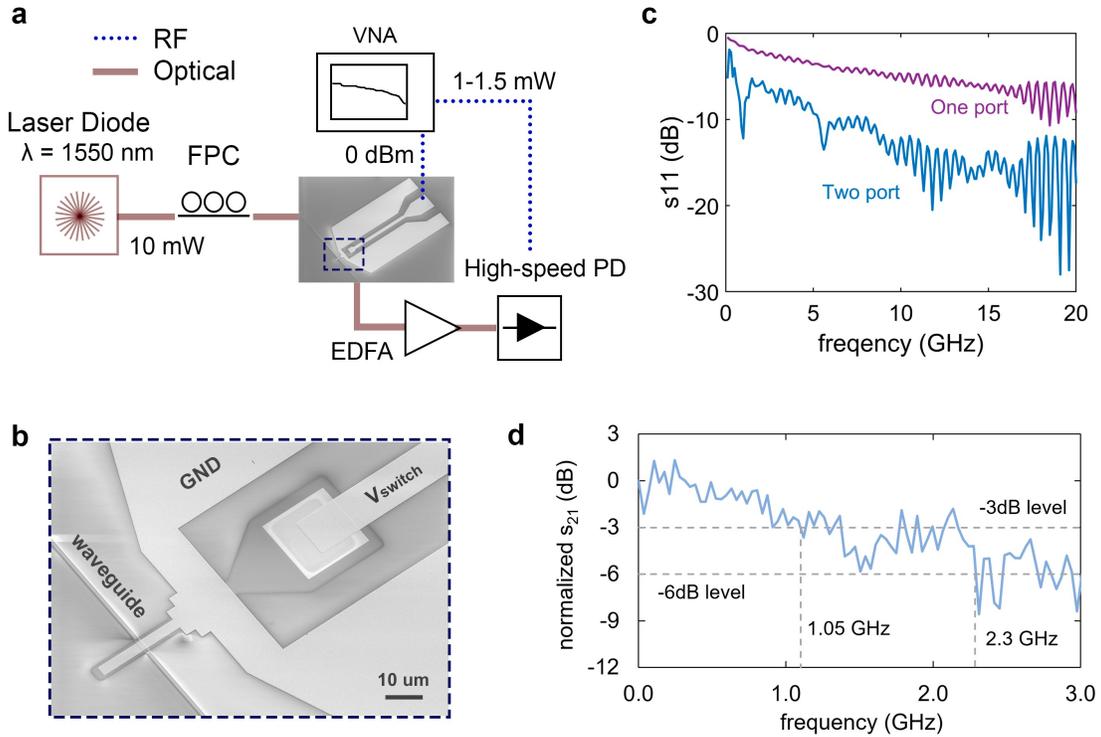

**Fig. 4** High-speed measurements: (a) Measurement setup; (b) SEM images of the fabricated RF electrodes; (c) $S_{11}$ and (d) $S_{21}$ frequency spectra.

To increase the extinction ratio and obtain the phase tunable, it is necessary to optimize the design. The top metal electrode must be changed to a transparent electrode, for example, the second ITO layer, as have been already proposed [15]. In the future it is planned to implement the technology developed in this work for the production of amplitude and phase designs with TE mode in silicon nitride platform.

## Conclusion

In conclusion, we have introduced the first ITO-based electro-optic modulator with absorption configuration integrated into $Si_3N_4$ PICs with the CMOS compatibility of the whole technology route. The devices demonstrate low insertion loss (5.7 dB at -7 V), sub-GHz electrical bandwidth (1.05 GHz), compact configuration (9.3 um) and extinction ratio of 1.5 dB. We worked with the waveguide TM mode to realize plasmonic enhancement of the mode/electro-optic material coupling.

## Methods

**ITO thin film deposition.** ITO thin films were deposited using ion beam assisted evaporation (IBAE) from pieces of ITO with a 99.999% purity ($In_2O_3$ - 90%, $SnO_2$ - 10%) at a base pressure below $3\times10^{-8}$ Torr, deposition rate of 2 Å/s and ion accelerating voltage of 70 V. Subsequent annealing was carried out in the 10% oxygen diluted argon atmosphere for 20 minutes to crystallize the film and activate oxygen.

**Device fabrication.** We first cleaned the silicon substrates in a 2:1 solution of sulfuric acid and hydrogen peroxide (80°C) and then rinsed with isopropanol to remove organic matter and immersed the plates in 49% hydrofluoric acid for about 20 seconds to remove the natural oxide layer. The wet oxidation processes at 1100°C was used to grow a 2.8-µm-thick thermal oxide layer. Then LPCVD process was performed at 800°C using $NH_3$ and $SiH_2Cl_2$ to form the $Si_3N_4$ waveguide layer. The Mach–Zehnder interferometers, grating couplers and tapers were patterned by electron-beam lithography (EBL) and reactive ion etching (process parameters can be found elsewhere [16]).

The ITO electrodes were fabricated using optical lift off lithography with a double resist mask. After ITO deposition, the mask was removed in N-Methyl-2-pyrrolidone solution. After lift off, the ITO structures were annealed.

$Al_2O_3$ was made by atomic layer deposition method. The process took place at a constant reactor temperature of 240 °C. Using trimethyl-aluminum (TMA) and $H_2O$ as the aluminum and oxygen precursors, respectively. Wet etching through the mask with antireflection layer was used to fabricate the contacts through $Al_2O_3$. It is known that $H_3PO_4$, MF-CD26, HF and TMAH are etchants, however, due to its selectivity to the photoresistive mask and the minimal lateral underetching, the $H_3PO_4$ solution was chosen. The $Al_2O_3$ etching rate was 3.3 nm/min. Finally, the Al electrodes were fabricated using electron beam lift off lithography [17].

**Modeling.** The effective mode indices and field distributions (Fig. 1a) in the both photonic and hybrid waveguide modes were calculated using 2D finite element method (FEM) simulation (COMSOL). The 800 nm thick PMLs were used as an open boundary condition. The real and imaginary parts of the ITO refractive index were calculated using the Drude-Lorentz model and entered directly into the ITO material parameters. For a more accurate calculation of the insertion loss and extinction coefficient, the ITO film was divided into two layers - 19 nm of the unchangeable layer and 1 nm of the accumulation layer, for which *n* and *k* were entered into the model. The maximum grid element size was 100 nm, the minimum was 0.25 nm with a narrow region resolution of 4.

The values of optical constants of the materials used in the modeling are given in Table 1. The ITO parameters used in the model are presented in Table 2.

**Table 1. Optical constants.**

| Parameter | Al | $Al_2O_3$ | ITO (at 0V) | $Si_3N_4$ | $SiO_2$ |
|---|---|---|---|---|---|
| n | 1.44 | 1.65 | 1.4707 | 2.00 | 1.445 |
| k | 16 | 0 | 0.1325 | 0 | 0 |

The Semiconductor module was used to estimate the change in carrier concentration in the layer. The boundary conditions of thin insulator gate, analytic doping model and metal contact were used.

The total losses and modulation depth were estimated using 3D FDTD (Ansys Optics) with an element mesh size of 0.5 nm in the z-direction in the plasmon propagation region. The total loss of the device consists of two components: the propagation loss and the coupling loss in the SPP. Knowing the total loss from 3D FDTD and the propagation loss from 2D FEM, the coupling loss can be determined as follows: $IL_{coupling}$ = $0.5(IL_{total} - IL_{propagation})$.

**Table 2. ITO electrical parameters.**

| | Relative permittivity | 9 |
|---|---|---|
| Semiconductor Analysis | Band gap, V | 3.8 |
| | Electron affinity, V | 5 |
| | Effective density of states, valence band, $cm^{-3}$ | $4 \cdot 10^{19}$ |
| | Effective density of states, conduction band, $cm^{-3}$ | $5 \cdot 10^{18}$ |
| | Electron mobility, $cm^2/(V \cdot s)$ | 17 |
| | Hole mobility, $cm^2/(V \cdot s)$ | 5 |

## Acknowledgements

Samples were fabricated and studied at the BMSTU Nanofabrication Facility (FMN Laboratory, FMNS REC, ID 74300).

## Additional information

The authors declare no conflict of interest.